\documentstyle[prl,aps,epsfig]{revtex}
\input epsf

\begin{document}
\title{Spin Dynamics and Orbital State in LaTiO$_{3}$}
\author{B. Keimer$^{1,2}$, D. Casa$^{2}$, A. Ivanov$^{3}$, J.W. Lynn$^{4}$, M. v.
Zimmermann$^{5}$,\\
J.P. Hill$^{5}$, D. Gibbs$^{5}$, Y. Taguchi$^{6}$, and Y. Tokura$^{6}$}
\address{$^1$ Max-Planck-Institut f\"{u}r Festk\"{o}rperforschung,
70569 Stuttgart, Germany\\
$^2$ Department of Physics, Princeton University, Princeton, NJ 08544\\
$^3$ Institut Laue-Langevin, 156X, 38042 Grenoble Cedex 9, France\\
$^4$ NIST Center for Neutron Research, National Institute of Standards and
Technology, Gaithersburg, MD 20899\\
$^5$ Department of Physics, Brookhaven National Laboratory, Upton, NY 11973\\
$^6$ Department of Applied Physics, University of Tokyo, Tokyo 113, Japan}
\date{\today}

\twocolumn[\hsize\textwidth\columnwidth\hsize\csname@twocolumnfalse\endcsname

\maketitle
           
\begin{abstract}
A neutron scattering study of the Mott-Hubbard insulator LaTiO$_{3}$ (T$_{{\rm N}}=132$ K) 
reveals a spin wave spectrum that is well described by a
nearest-neighbor superexchange constant $J=15.5$ meV and a small
Dzyaloshinskii-Moriya interaction ($D=1.1$ meV). The nearly isotropic spin
wave spectrum is surprising in view of the absence of a static Jahn-Teller
distortion that could quench the orbital angular momentum, 
and it may indicate strong orbital fluctuations. A resonant x-ray scattering study
has uncovered no evidence of orbital order in LaTiO$_{3}$.
\end{abstract}


]

In the layered cuprates exemplified by the series La$_{2-x}$Sr$_{x}$ CuO$%
_{4+\delta }$, the transition from a $3d^{9}$ antiferromagnetic (AF)
insulator at $x=\delta =0$ into an unconventional metallic and
superconducting state with increasing hole concentration ($x,$ $\delta >0)$
has received an enormous amount of attention. The magnetic spectra of these
materials, revealed by inelastic neutron scattering, have played a key role
in efforts to arrive at a theoretical explanation of this transition. The
pseudocubic perovskite La$_{1-x}$Sr$_{x}$TiO$_{3+\delta}$ undergoes an
analogous transition from a $3d^{1}$ AF insulator at $x=\delta =0$ to a
metallic state with increasing hole concentration \cite{tokura93}. In the
titanates, however, the metallic state shows conventional Fermi liquid
behavior, and no superconductivity is found \cite{tokura93}.
Momentum-resolved probes such as angle-resolved photoemission spectroscopy
and inelastic neutron scattering have thus far not been applied to the
titanates, and the origin of the very different behavior of the metallic
cuprates and titanates is still largely unexplored. Here we report an
inelastic neutron scattering and anomalous x-ray scattering study of the
parent compound of the titanate series, LaTiO$_{3}$, that provides insight
into the microscopic interactions underlying this behavior.

Orbital degrees of freedom, quenched in the layered cuprates by a large
Jahn-Teller (JT) distortion of the CuO$_{6}$ octahedra, are likely to be a
key factor in the phenomenology of the titanates. While the TiO$_{6}$
octahedra are {\it tilted} in a GdFeO$_{3}$-type structure, their {\it %
distortion} is small and essentially undetectable in neutron powder
diffraction experiments on LaTiO$_{3}$ (Ref. \cite{greedan79}). The crystal
field acting on the Ti$^{3+}$ ion is therefore nearly cubic, and
heuristically one expects a quadruply degenerate single-ion ground state
with unquenched orbital angular momentum opposite to the spin angular
momentum due to the spin-orbit interaction. In other perovskites such as
LaMnO$_3$, such spin-orbital degeneracies are broken by successive orbital
and magnetic ordering transitions \cite{murakami98}. In the orbitally and
magnetically ordered state of LaMnO$_{3}$, the spin wave spectrum is highly
anisotropic reflecting the different relative orientations of the orbitals
on nearest-neighbor Mn atoms in different crystallographic directions \cite
{moussa96}.

The reduced ordered moment ($\mu _{0}\sim 0.45\mu _{B}$, Ref. \cite
{greedan83}) in the G-Type AF structure of LaTiO$_{3}$ (inset in Fig.~\ref
{fig1}) at first sight appears consistent with a conventional scenario in
which the orbital occupancies at every site are established at some high
temperature, and the magnetic degrees of freedom (coupled spin and orbital
angular momenta) order at a lower temperature. Full theoretical
calculations, however, generally predict a ferromagnetic spin structure for
LaTiO$_{3}$ \cite{fujimori96}. As in LaMnO$_{3}$, the spin dynamics of LaTiO$%
_{3}$ are highly sensitive to the orbital occupancies and can provide
important information in this regard. We find that the exchange anisotropy
is small and hence inconsistent with the presence of an appreciable
unquenched orbital moment. At the same time, synchrotron x-ray scattering
experiments have not revealed any evidence of reflections showing a resonant
enhancement at the Ti K-edge, unlike other perovskites in which orbital
order (OO) is present. These observations, along with previously puzzling
Raman scattering data \cite{reedyk97}, indicate strong fluctuations in the
orbital sector of LaTiO$_{3}$.

The neutron scattering experiments were conducted on the BT2 and BT4 triple
axis spectrometers at the NIST research reactor and at the IN8 spectrometer
at the Institut Laue-Langevin. For excitation energies up to 25 meV, we used
high resolution configurations with vertically focusing pyrolytic graphite
(PG) (002) monochromator and PG (002) analyser crystals set for final
neutron energies of 14.7 meV or 30.5 meV and horizontally collimated beams
at both NIST and ILL. For excitation energies of 20 meV and higher, we used
a double-focusing analyser on IN8 with open collimations, and with a Cu
(111) monochromator and a PG (002) analyser set for a final neutron energy
of 35 meV. PG filters were used in the scattered beam to reduce higher order
contamination. Data obtained on the different spectrometers and with
different configurations were in good agreement.

The sample was a single crystal of volume 0.2 cm$^{3}$ and mosaicity 0.5$%
^{\circ}$ grown by the floating zone technique. It is semiconducting, and
neutron diffraction (Fig.~\ref{fig1}) shows a sharp, second-order N\'{e}el
transition at T$_{{\rm N}}=132$ K, implying a highly homogeneous oxygen
content ($\delta \sim 0.01$ in LaTiO$_{3+\delta}$) \cite{tokura93}. The
diffraction pattern is consistent with the G-type structure found previously 
\cite{greedan83}, and a small uncompensated moment ($\sim 10^{-2}\mu _{B}$
per Ti spin at low temperatures) appears below T$_{{\rm N}}$ due to spin
canting, as observed by magnetization measurements. The
nuclear structure of LaTiO$_{3}$ is orthorhombic (space group Pnma \cite
{greedan79}), but the crystal is fully twinned. Because of the isotropy of the
spin wave dispersions (see below), twinning did not influence the
neutron measurements. For simplicity we express the
wave vectors in the pseudocubic notation with lattice constant $a\sim 3.95$%
\AA. In this notation, AF Bragg reflections are located at ($h/2,k/2,l/2$)
with $h,k,l$ odd. Data were taken with the crystal in two different
orientations in which wave vectors of the form ($h,h,l$) or ($h,k,(h+k)/2$),
respectively, were accessible.

Fig.~\ref{fig2} shows inelastic neutron scattering data obtained in constant-%
{\bf q} mode with high resolution near the AF zone center. The dispersing
peak shown disappears above the N\'{e}el temperature, thus clearly
identifying itself as a spin wave excitation. In Fig.~\ref{fig3},
constant-energy scans obtained in the high-intensity configuration with
relaxed resolution are presented. The profile shapes are strongly influenced
by the spectrometer resolution, and a deconvolution is required to
accurately extract the positions of the spin wave peaks. For the
high-resolution configuration with only vertical focusing we used the
standard Cooper-Nathans procedure while a Monte-Carlo ray-tracing routine
was used for the doubly focused geometry \cite{kulda}. An analytical
approximation to the Ti$^{3+}$ form factor and the standard intensity
factors for AF magnons were incorporated in the programs \cite{wilson}. The
peak positions thus obtained are shown in Fig.~\ref{fig4} in the (111)
direction.

A very good global fit to all data was obtained by convoluting the
resolution function with a single spin wave branch of the generic form 
%TCIMACRO{\UNICODE{0x127}}%
%BeginExpansion
h\hskip-.2em\llap{\protect\rule[1.1ex]{.325em}{.1ex}}\hskip.2em%
%EndExpansion
$\omega =zSJ\sqrt{(1+\epsilon )^{2}-\gamma ^{2}\text{ }}$where 
%TCIMACRO{\UNICODE{0x127}}%
%BeginExpansion
h\hskip-.2em\llap{\protect\rule[1.1ex]{.325em}{.1ex}}\hskip.2em%
%EndExpansion
$\omega $ is the spin wave energy, $z=6$ is the coordination number, $S=1/2$
is the Ti spin, $J=15.5\pm 1$ meV is the isotropic (Heisenberg) part of the
nearest-neighbor superexchange \cite{note}, $\gamma =\frac{1}{3}[\cos
(q_{x}a)+\cos (q_{y}a)+\cos (q_{z}a)]$ with the magnon wave vector {\bf q}
measured from the magnetic zone center, and the zone center gap is $\Delta
\sim zSJ\sqrt{2\epsilon }=3.3\pm 0.3$ meV. The solid lines in Figs.~\ref
{fig2}-\ref{fig4} result from this global fit and obviously provide a good
description of all data. Inclusion of further-neighbor interactions, damping 
parameters above the instrumental resolution, or other
(nondegenerate) spin wave branches did not improve the fit.

The Heisenberg exchange constant $J$ is in fair agreement with predictions
based on a comparison of the N\'{e}el temperature with numerical simulations
(T$_{{\rm N}}=0.946J/k_{B}\sim 170$ K for spins-1/2 on a simple cubic
lattice \cite{sandvik99}). In general, the spin wave gap is determined by
symmetric and antisymmetric (Dzyaloshinskii-Moriya) anisotropy terms in the
superexchange matrix, by terms originating from direct exchange, by dipolar
interactions, and by the single ion anisotropy. The latter two effects are
negligible and nonexistent, respectively, in spin-1/2 systems. In the GdFeO$%
_{3}$ structure, antisymmetric exchange is allowed by symmetry, and its
magnitude is expected to scale with the tilt angle of the ${\rm TiO_{6}}$
octahedra. Because of the large tilt angle ($11.5^{\circ }$), we expect this
effect to dominate over the more subtle direct exchange terms. Theories of
superexchange anisotropies \cite{moriya60} were recently reexamined \cite
{aharony95} in the light of neutron scattering data on the layered cuprates.
It was shown that the symmetric and antisymmetric terms are related by a
hidden symmetry so that the two zone-center spin wave gaps depend on the
relationship between Dzyaloshinkii-Moriya (DM) vectors centered on each
magnetic bond. Specifically, for two-dimensional (2D) spin structures (such
as the one of La$_{2}$CuO$_{4}$) the gaps are degenerate if all DM vectors
have the same magnitude, and the degenerate gaps are nonzero if, in
addition, not all vectors have the same orientation. The bond-dependent DM
vectors for LaMnO$_{3}$ (isostructural to LaTiO$_{3}$) were given in Ref. 
\cite{solovyev96}. Although a detailed analysis of the spin dynamics for
these 3D systems has not been reported, we note that both of the above
criteria are fulfilled for LaTiO$_{3}$. In particular, the DM vectors
centered on the six Ti-O-Ti bonds have different orientations but are
expected to have the same magnitude because of the practically equal bond
lengths and bond angles \cite{greedan79}. Our observation of degenerate but
nonzero spin wave gaps therefore provides support for the predictions of
Ref. \cite{aharony95} in a 3D, non-cuprate system.

Carrying the analogy to the 2D cuprates one step further, we expect that the
anisotropy gap is $\Delta \sim zSD$, and hence $D\sim 1.1$ meV is the net DM
interaction per Ti spin. Due to spin canting, the net ferromagnetic moment
per spin should therefore be $\sim \mu _{0}D/2J=1.5\times 10^{-2}\mu _{B}$,
in good agreement with the observed value. This supports our assumption that
the DM interaction provides the dominant contribution to the spin wave gap.
A microscopic calculation of $D$ would be a further interesting test of the
formalism developed in Ref. \cite{aharony95}.

The small easy-axis anisotropy in the exchange Hamiltonian of LaTiO$_{3}$ is
difficult to understand based on simple crystal-field considerations for the
Ti$^{3+}$ ion. The antisymmetric exchange is generally of order $(\Delta
g/g)J$ where $g$ is the free-electron Land\'{e} factor and $\Delta g$ is its
shift in the crystalline environment \cite{moriya60}. Based on our data, we
therefore estimate $\Delta g/g\sim 0.05.$ On the other hand, in the absence
of any appreciable static JT distortion as observed experimentally \cite
{greedan79}, one expects that the spin-orbit interaction ( $\Lambda \sim +20$
meV \cite{abragam}) splits the $t_{2g}$ multiplet of the cubic crystal field
Hamiltonian into a quadruply degenerate ground state and a higher-lying
Kramers doublet. In a simple crystal field model, this ground state is
characterized by an unquenched orbital moment equal and antiparallel to the
spin moment ($\Delta g/g=1$). More elaborate Hartree-Fock calculations \cite
{fujimori96} do not change this picture qualitatively: Even if a static JT
distortion at the limits of the experimental error bars \cite{greedan79} is
included, the orbital contribution to the moment remains comparable to the
spin moment ($\Delta g/g\sim 0.5$). There is thus an order-of-magnitude
discrepancy between the predictions of conventional models and the neutron
scattering obervations. The smallness of the spin anisotropy in LaTiO$_{3}$
is underscored by a different comparison: The $D/J$ ratio of LaTiO$%
_{3}$ differs only by a factor of 3 from that of La$_{2}$CuO$_{4}$, whose
low-temperature ordered moment is in near-perfect agreement with the
spin-only prediction. Since $D/J$ scales with the tilt angle of the
octahedra which is a factor of $\sim 3$ larger in LaTiO$_{3}$, the relative
magnitude of this quantity in the two materials can be accounted for without
invoking a large orbital moment in LaTiO$_{3}$.

Interestingly, the large discrepancy between the predictions of conventional
models and the neutron scattering observations has a close analogy in the
electron spin resonance (ESR) literature. The description of ESR data on Ti$%
^{3+}$ impurities embedded into perovskite lattices in fact commonly
requires $g$-factors that are much more isotropic than predicted by simple
crystal field calculations \cite{abragam}. According to a widely used model 
\cite{ham65}, this is attributed to the dynamical JT effect where the
orbital degeneracy is lifted by coupling to zero-point lattice vibrations.

While the dynamical JT effect is well established in impurity systems, it
has thus far not been reported in lattice systems which commonly exhibit
static, cooperative JT distortions associated with OO. Anomalous x-ray
diffraction with photon energies near an absorption edge of the transition
metal ion has recently been established as a direct probe of OO in
perovskites \cite{murakami98} whose sensitivity far exceeds conventional
diffraction techniques that probe OO indirectly through associated lattice
distortions. We have therefore carried out an extensive search for
reflections characteristic of orbital ordering near the Ti K-edge (4.966
keV) at beamline X22C at the National Synchrotron Light Source, with energy
resolution $\sim 5$ eV. The experiments were performed at low temperature
(T=10K) on a polished (1,1,0) surface of the same crystal that was also used
for the neutron measurements. No evidence for resonant reflections at
several high symmetry positions (such as ($\frac12$,$\frac12$,0), the
ordering wave vector expected for $t_{2g}$ orbitals with a G-type spin
structure \cite{ren98}) was found under the same conditions that enabled
their positive identification in LaMnO$_{3}$ \cite{murakami98}, YTiO$_{3}$ 
\cite{ytio3} and related materials. If OO is present in LaTiO$_{3}$, 
we can hence conclude that its
order parameter is much smaller than in comparable perovskites. On general
grounds, the reduction of the order parameter should be accompanied by
enhanced orbital zero-point fluctuations. These may have already been
detected (though not identified as such): A large electronic background and
pronounced Fano-type phonon anomalies were observed by Raman scattering in
nominally stoichiometric, {\it insulating} titanates and are most pronounced in
LaTiO$_{3}$ \cite{reedyk97}. In the light of our observations, it is of
course important to revisit these experiments and rule out any possible role
of residual oxygen defects, inhomogeneity, etc. The presently available data
are, however, naturally interpreted as arising from orbital fluctuations
coupled to lattice vibrations.

An observation not explained by these qualitative considerations is the 
small ordered moment in the AF state. If the orbital moment is indeed 
largely quenched, one would naively expect a spin-only moment of $0.85 \mu _{B}$ 
\cite{anderson52}, in contrast to the experimental observation of $0.45 \mu_{B}$. 
A recently proposed full theory of the interplay between the orbital and spin 
dynamics in LaTiO$_3$ has yielded a prediction of the ordered moment that is 
in quantitative agreement with experiment \cite{khaliullin}.

In conclusion, several lines of evidence from neutron, x-ray and Raman
scattering can be self-consistently interpreted in terms of an unusual many
body state with AF long range order but strong orbital fluctuations. This
should be an interesting subject of theoretical research. The orbital
fluctuations are expected to be enhanced in the presence of itinerant charge
carriers and therefore to strongly influence the character of the
insulator-metal transition. The present study provides a starting point for
further investigations in doped titanates.

We thank A. Aharony, M. Cardona, P. Horsch, D. Khomskii, E.
M\"{u}ller-Hartmann, A. Oles, G. Sawatzky, and especially G. Khaliullin for
discussions, and J. Kulda for gracious assistance with his resolution
program. The work was supported by the US-NSF under grant No. DMR-9701991,
by the US-DOE under contrat No. DE-AC02-98CH10886, and by NEDO and
Grants-In-Aid from the Ministry of Education, Japan.

\begin{figure}
\caption{
Integrated intensity of the (0.5,0.5,0.5) AF Bragg reflection as a
function of temperature. The inset shows the G-type spin structure.
}
\label{fig1}
\end{figure}

\begin{figure}
\caption{
Typical constant-{\bf q} scans near the AF zone center. The upper
profile is offset by 200 units. The lines are the result of a convolution of
a spin wave cross section with the instrumental resolution function, as
described in the text.
}
\label{fig2}
\end{figure}

\begin{figure}
\caption{
Typical constant-energy scans in two different directions of
reciprocal space. The profiles are labeled by the excitation energy in meV.
The lines are the result of a convolution of a spin wave cross section with
the instrumental resolution function, as described in the text.
}
\label{fig3}
\end{figure}

\begin{figure}
\caption{
Fitted spin wave peak positions in the (1,1,1) direction of
reciprocal space. The line is the magnon dispersion curve described in the
text.
}
\label{fig4}
\end{figure}


\begin{references}
\bibitem{note} $J$ is the superexchange energy per nearest-neighbor pair.
In this notation $J\sim 100$ meV in La$_{2}$CuO$_{4}$.
\end{references}
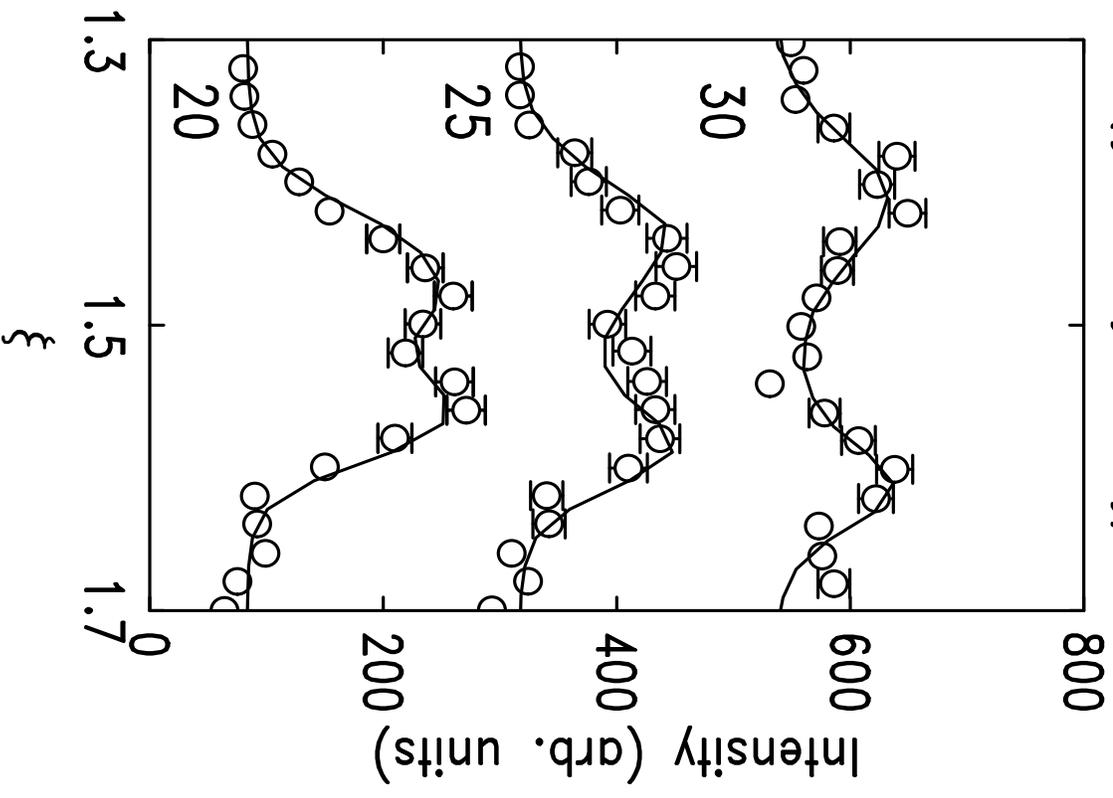
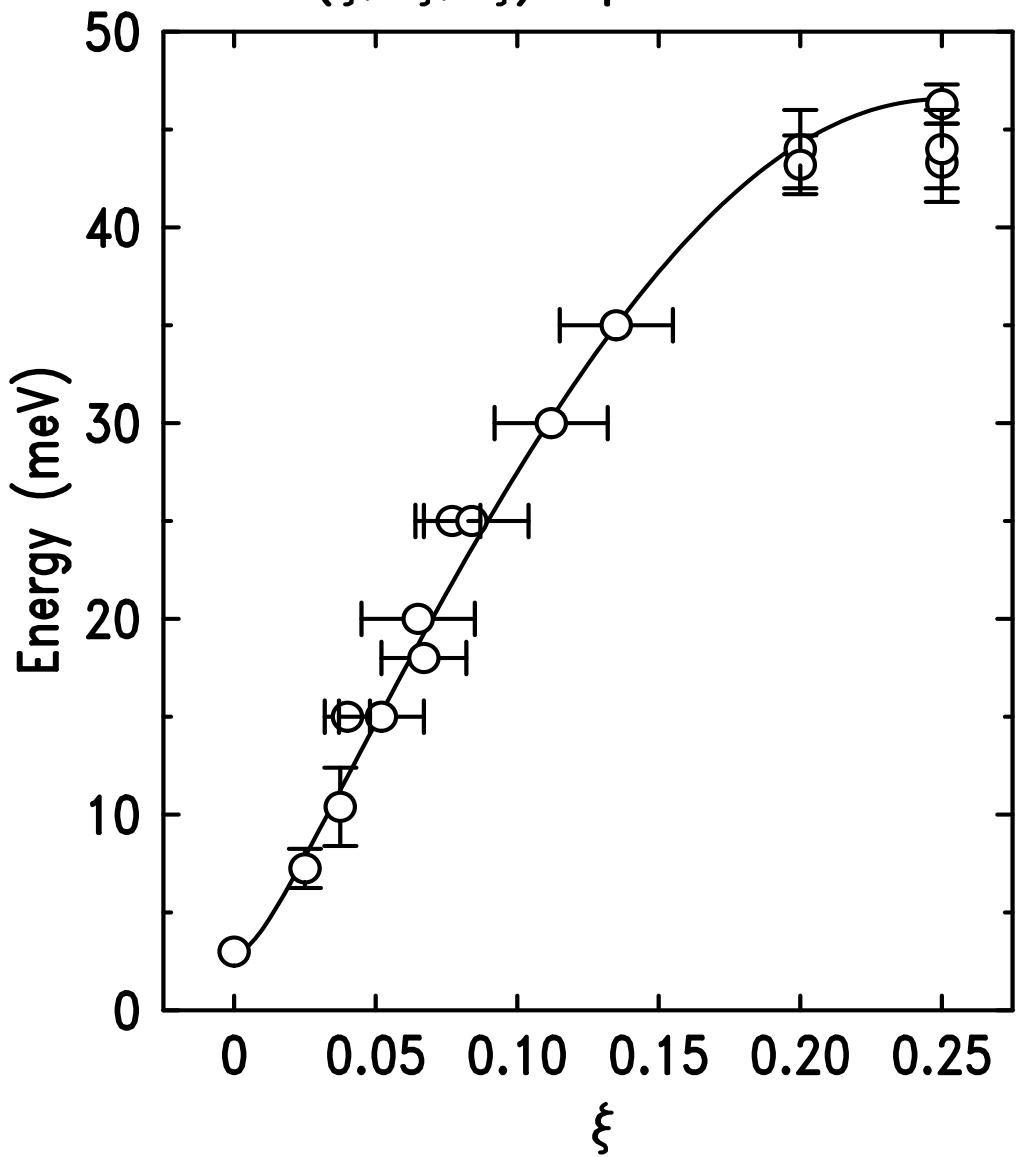
\end{document}